\documentstyle[11pt,paspconf,psfig]{article}

\def\gsim{${\raise.3ex\hbox{$>$\kern-.75em\lower1ex\hbox{$\sim$}}}$}
\def\lsim{${\raise.3ex\hbox{$<$\kern-.75em\lower1ex\hbox{$\sim$}}}$}
\def\lcdm{$\Lambda$CDM}

\begin{document}

\title{STATUS OF COSMOLOGY}

\author{Joel R. Primack}

\affil{University of California, Santa Cruz, CA 95064 U.S.A.}

\begin{abstract}
The cosmological parameters that I will emphasize are the Hubble
parameter $H_0 \equiv 100 h$ km s$^{-1}$ Mpc$^{-1}$, the age of the
universe $t_0$, the average matter density $\Omega_m$, the baryonic
matter density $\Omega_b$, the neutrino density $\Omega_\nu$, and the
cosmological constant $\Omega_\Lambda$. The evidence currently favors
$t_0 \approx 13$ Gyr, $h \approx 0.65$, $\Omega_m \approx 0.3$,
$\Omega_\Lambda \approx 0.7$.
\end{abstract}

\section{Introduction}

In this brief summary I will concentrate on the values of the
cosmological parameters. The other key questions in
cosmology today concern the nature of the dark matter
and dark energy, the origin and nature of the primordial
inhomogeneities, and the formation and evolution of galaxies.  I have
been telling my theoretical cosmology students for several years that
these latter topics are their main subjects for research, since
determining the values of the cosmological parameters is now mainly in
the hands of the observers.

In discussing cosmological parameters, it will be useful to
distinguish between two sets of assumptions: (a) general relativity
plus the assumption that the universe is homogeneous and isotropic on
large scales (Friedmann-Robertson-Walker framework), or (b) the \lcdm\
family of models.  The \lcdm\ models assume that the present matter
density $\Omega_m$ plus the cosmological constant (or its equivalent
in ``dark energy'') in units of critical density $\Omega_\Lambda =
\Lambda/(3 H_0^2)$ sum to unity ($\Omega_m + \Omega_\Lambda = 1$) to
produce the flat universe predicted by simple cosmic inflation models.
The \lcdm\ family of models was introduced by Blumenthal et
al. (1984), who worked out the linear power spectra $P(k)$ and a
semi-analytic treatment of structure formation compared to the
then-available data.  We did ths for the $\Omega_m=1$, $\Lambda=0$
``standard'' cold dark matter (CDM) model, and also for the
$\Omega_m=0.2$, $\Omega_\Lambda=0.8$ \lcdm\ model.  In addition to
$\Omega_m + \Omega_\Lambda = 1$, these \lcdm\ models assumed that the
primordial fluctuations were Gaussian with a Zel'dovich spectrum
($P_p(k)=Ak^n$, with $n=1$), and that the dark matter is mostly of the
cold variety.

The  table below summarizes the current observational information
about the cosmological parameters.  The quantities in brackets have
been deduced using at least some of the \lcdm\ assumptions.  The rest
of this paper discusses these issues in more detail.  But it should
already be apparent that there is impressive agreement between the
values of the parameters determined by various methods.

\begin{table}
\caption{Cosmological Parameters [results assuming \lcdm\ in
brackets]}
\label{ta:parameters}
\centerline{\vbox{\halign{\ \ #\hfill \quad \qquad &$#$\hfill \ &$#$
&#\hfill \ \ \cr
\noalign{\hrule}
\noalign{\vskip .10in}
Hubble parameter  &H_0        &= &$100 \,h$ km s$^{-1}$ Mpc$^{-1}$ ,
                                \ $h = 0.65 \pm 0.08$ \cr
Age of universe   &t_0        &= &9-16 Gyr (from globular clusters) \cr
{}              &{}           &= &[9-17 Gyr] \cr
Baryon density  &\Omega_b h^2 &= &$0.019\pm0.001$ (from D/H) \cr
{}              &{}           &> &[0.015 from Ly$\alpha$ forest opacity] \cr
Matter density  &\Omega_m     &= &$0.4\pm0.2$ (from cluster baryons) \cr
{}              &{}             &= & [$0.34\pm0.1$ from Ly$\alpha$
forest $P(k)$] \cr
{}              &{}             &= & [$0.4\pm0.2$ from cluster evolution] \cr
{}              &{}           &> & $0.3$ ($2.4 \sigma$, from flows) \cr
{}              &{}    &\approx &${3\over4} \Omega_\Lambda - {1\over4}
                        \pm{1\over8}$ from SN Ia \cr
Total density   &\Omega_m + \Omega_\Lambda  &\approx &$1\pm0.3$
                       (from CMB peak location) \cr
Dark energy density &\Omega_\Lambda &= &$0.8 \pm 0.3$ (from previous
                      two lines) \cr
{}              &{}    &< & 0.73 (2$\sigma$) from radio QSO lensing \cr
Neutrino density &\Omega_\nu  &$\gsim$ &0.001 (from Superkamiokande) \cr
{}              &{}           &$\lsim$ &[0.1] \cr
\noalign{\vskip .10in}
\noalign{\hrule}
}}}
\end{table}

\section{Age of the Universe $t_0$}

The strongest lower limits for $t_0$ come from studies of the stellar
populations of globular clusters (GCs).  In the mid-1990s the best
estimates of the ages of the oldest GCs from main sequence turnoff
magnitudes were $t_{GC} \approx 15-16$ Gyr (Bolte \& Hogan 1995;
VandenBerg, Bolte, \& Stetson 1996; Chaboyer et al. 1996).  A
frequently quoted lower limit on the age of GCs was $12$ Gyr (Chaboyer
et al. 1996), which was then an even more conservative lower limit on
$t_0 = t_{GC} + \Delta t_{GC}$, where $\Delta t_{GC} \gsim 0.5$ Gyr is
the time from the Big Bang until GC formation. The main uncertainty in
the GC age estimates came from the uncertain distance to the GCs: a
0.25 magnitude error in the distance modulus translates to a 22\%
error in the derived cluster age (Chaboyer 1995).

In spring of 1997, analyses of data from the Hipparcos astrometric
satellite indicated that the distances to GCs assumed in obtaining the
ages just discussed were systematically underestimated (Reid 1997,
Gratton et al. 1997).  It follows that their stars at the main
sequence turnoff are brighter and therefore younger.  Stellar
evolution calculation improvements also lowered the GC age estimates.
In light of the new Hipparcos data, Chaboyer et al. (1998) have done a
new Monte Carlo analysis of the effects of varying various uncertain
parameters, and obtained $t_{GC} = 11.5 \pm 1.3$ Gyr ($1\sigma$), with
a 95\% C.L.  lower limit of 9.5 Gyr.  The latest detailed analysis
(Carretta et al.  1999) gives $t_{GC} = 11.8\pm2.6$ Gyr from main
sequence fitting using parallaxes of local subdwarfs, the method used
in the 1997 analyses quoted above. These authors get somewhat smaller
GC distances when all the available data is used, with a resulting
$t_{GC} = 13.2 \pm 2.9$ Gyr (95\% C.L.).

Stellar age estimates are of course based on standard stellar
evolution calculations.  But the solar neutrino problem reminds us
that we are not really sure that we understand how even our nearest
star operates; and the sun plays an important role in calibrating
stellar evolution, since it is the only star whose age we know
independently (from radioactive dating of early solar system
material).  An important check on stellar ages can come from
observations of white dwarfs in globular and open clusters (Richer et
al. 1998).

What if the GC age estimates are wrong for some unknown reason? The
only other non-cosmological estimates of the age of the universe come
from nuclear cosmochronometry --- radioactive decay and chemical
evolution of the Galaxy --- and white dwarf cooling. Cosmochronometry
age estimates are sensitive to a number of uncertain issues such as
the formation history of the disk and its stars, and possible actinide
destruction in stars (Malaney, Mathews, \& Dearborn 1989; Mathews \&
Schramm 1993). However, an independent cosmochronometry age estimate
of $15.6\pm4.6$ Gyr has been obtained based on data from two
low-metallicity stars, using the measured radioactive depletion of
thorium (whose half-life is 14.2 Gyr) compared to stable heavy
r-process elements (Cowan et al. 1997, 1999).  This method could
become very important if it were possible to obtain accurate
measurements of r-process element abundances for a number of very low
metallicity stars giving consistent age estimates, and especially if
the large errors could be reduced.

Independent age estimates come from the cooling of white dwarfs in the
neighborhood of the sun. The key observation is that there is a lower
limit to the luminosity, and therefore also the temperature, of nearby
white dwarfs; although dimmer ones could have been seen, none have
been found (cf. however Harris et al. 1999).  The only plausible
explanation is that the white dwarfs have not had sufficient time to
cool to lower temperatures, which initially led to an estimate of
$9.3\pm2$ Gyr for the age of the Galactic disk (Winget et al. 1987).
Since there was evidence, based on the pre-Hipparcos GC distances, 
that the stellar disk of our Galaxy is about 2 Gyr younger than the
oldest GCs (e.g., Stetson, VandenBerg, \& Bolte 1996, Rosenberg et
al. 1999), this in turn gave an estimate of the age of the universe of
$t_0 \approx 11\pm2$ Gyr. Other analyses (cf. Wood 1992, Hernanz et
al. 1994) conclude that sensitivity to disk star formation history,
and to effects on the white dwarf cooling rates due to C/O separation
at crystallization and possible presence of trace elements such as
$^{22}$Ne, allow a rather wide range of ages for the disk of about
$10\pm4$ Gyr. One determination of the white dwarf luminosity
function, using white dwarfs in proper motion binaries, leads to a
somewhat lower minimum luminosity and therefore a somewhat higher
estimate of the age of the disk of $\sim 10.5^{+2.5}_{-1.5}$ Gyr
(Oswalt et al. 1996; cf. Chabrier 1997).  More recent observations
(Leggett, Ruiz and Bergeron 1998) and analyses (Benvenuto \& Althaus
1999) lead to an estimated age of the galactic disk of $8 \pm 1.5$
Gyr.

We conclude that $t_0 \approx 13$ Gyr, with $\sim 11$ Gyr a lower
limit. Note that $t_0 > 13$ Gyr implies that $h \leq 0.50$ for matter
density $\Omega_m=1$, and that $h \leq 0.73$ even for $\Omega_m $ as
small as 0.3 in flat cosmologies (i.e., with $\Omega_m +
\Omega_\Lambda = 1$). If $t_0$ is as low as $\sim 11$ Gyr, that would
allow $h$ as high as 0.6 for $\Omega_m=1$.

\section{Hubble Parameter $H_0$}

The Hubble parameter $H_0\equiv 100 h$ km s$^{-1}$ Mpc$^{-1}$ remains
uncertain, although no longer by the traditional factor of two.  The
range of $h$ determinations has been shrinking with time (Kennicutt,
Freedman, \& Mould 1995).  De~Vaucouleurs long contended that $h
\approx 1$. Sandage has long contended that $h \approx 0.5$, although
a recent reanalysis of the Type Ia supernovae (SNe Ia) data coauthored by
Sandage and Tammann concludes that the latest data are consistent with
$h=0.6\pm0.04$ (Saha et al. 1999).

The Hubble parameter has been measured in two basic ways: (1)
Measuring the distance to some nearby galaxies, typically by measuring
the periods and luminosities of Cepheid variables in them; and then
using these ``calibrator galaxies'' to set the zero point in any of
the several methods of measuring the relative distances to
galaxies. (2) Using fundamental physics to measure the distance to
some distant object(s) directly, thereby avoiding at least some of the
uncertainties of the cosmic distance ladder (Rowan-Robinson 1985). The
difficulty with method (1) was that there was only a handful of
calibrator galaxies close enough for Cepheids to be resolved in
them. However, the HST Key Project on the Extragalactic Distance Scale
has significantly increased the set of calibrator galaxies.  The
difficulty with method (2) is that in every case studied so far, some
aspect of the observed system or the underlying physics remains
somewhat uncertain. It is nevertheless remarkable that the results of
several different methods of type (2) are rather similar, and indeed
not very far from those of method (1). This gives reason to hope for
convergence.

\subsection{Relative Distance Methods}

One piece of good news is that the several methods of measuring the
relative distances to galaxies now mostly seem to be consistent with
each other. These methods use either ``standard candles'' or empirical
relations between two measurable properties of a galaxy, one
distance-independent and the other distance-dependent. The favorite
standard candle is SNe Ia, and observers are now
in good agreement.  Taking account of an empirical relationship
between the SNe Ia light curve shape and maximum luminosity leads to $h
= 0.65\pm0.06$ (Riess, Press, \& Kirshner 1996),
$h=0.64^{+0.08}_{-0.06}$ (Jha et al. 1999), or $h = 0.63\pm0.03$
(Hamuy et al. 1996, Phillips et al. 1999), and the slightly lower
value mentioned above from the latest analysis coauthored by Sandage
and Tammann agrees within the errors.  The HST Key Project result
using SNe Ia is $h = 0.65 \pm 0.02 \pm 0.05$, where the first error
quoted is statistical and the second is systematic (Gibson et al. 1999),
and their luminosity-metallicity relationship (Kennicutt et al. 1998)
has been used (this lowers $h$ by 4\%).  Some of the other relative
distance methods are based on old stellar populations: the tip of the
red giant branch (TRGB), the planetary nebula luminosity function
(PNLF), the globular cluster luminosity function (GCLF), and the
surface brightness fluctuation method (SBF). The HST Key Project
result using these old star standard candles is $h=0.66\pm 0.04 \pm
0.06$. The old favorite empirical relation used as a relative distance
indicator is the Tully-Fisher relation between the rotation velocity
and luminosity of spiral galaxies.  The ``final'' value of the Hubble
constant from the HST Key Project taking all of these into account is
$h=0.71\pm0.06$ (Ferrarese et al. 1999, and this conference, for a
nice summary).

\subsection{Fundamental Physics Approaches}

The fundamental physics approaches involve either Type Ia or Type II
supernovae, the Sunyaev-Zel'dovich (S-Z) effect, or gravitational
lensing of quasars.  All are promising, but in each case the relevant
physics remains somewhat uncertain.

The $^{56}$Ni radioactivity method for determining $H_0$ using Type Ia
SNe avoids the uncertainties of the distance ladder by calculating the
absolute luminosity of Type Ia supernovae from first principles using
plausible but as yet unproved physical models for $^{56}$Ni
production.  The first result obtained was that $h=0.61\pm0.10$
(Arnet, Branch, \& Wheeler 1985; Branch 1992); however, another study
(Leibundgut \& Pinto 1992; cf.  Vaughn et al. 1995) found that
uncertainties in extinction (i.e., light absorption) toward each
supernova increases the range of allowed $h$. Demanding that the
$^{56}$Ni radioactivity method agree with an expanding photosphere
approach leads to $h=0.60^{+0.14}_{-0.11}$ (Nugent et al. 1995). The
expanding photosphere method compares the expansion rate of the SN
envelope measured by redshift with its size increase inferred from its
temperature and magnitude.  This approach was first applied to Type II
SNe; the 1992 result $h=0.6\pm0.1$ (Schmidt, Kirschner, \& Eastman
1992) was subsequently revised upward by the same authors to
$h=0.73\pm0.06\pm0.07$ (1994). However, there are various
complications with the physics of the expanding envelope
(Ruiz-Lapuente et al. 1995; Eastman, Schmidt, \& Kirshner 1996).

The S-Z effect is the Compton scattering of microwave background
photons from the hot electrons in a foreground galaxy cluster.  This
can be used to measure $H_0$ since properties of the cluster gas
measured via the S-Z effect and from X-ray observations have different
dependences on $H_0$.  The result from the first cluster for which
sufficiently detailed data was available, A665 (at $z=0.182$), was
$h=(0.4-0.5)\pm0.12$ (Birkinshaw, Hughes, \& Arnoud 1991); combining
this with data on A2218 ($z=0.171$) raised this somewhat to
$h=0.55\pm0.17$ (Birkinshaw \& Hughes 1994).  The history and more
recent data have been reviewed by Birkinshaw (1999), who concludes
that the available data give a Hubble parameter $h\approx0.6$ with a
scatter of about 0.2.  But since the available measurements are not
independent, it does not follow that $h=0.6\pm0.1$; for example, there
is a selection effect that biases low the $h$ determined this way.

Several quasars have been observed to have multiple images separated
by $\theta \sim$ a few arc seconds; this phenomenon is interpreted as
arising from gravitational lensing of the source quasar by a galaxy
along the line of sight (first suggested by Refsdal 1964; reviewed in
Williams \& Schechter 1997).  In the first such system discovered, QSO
0957+561 ($z=1.41$), the time delay $\Delta t$ between arrival at the
earth of variations in the quasar's luminosity in the two images has
been measured to be, e.g., $409\pm23$ days (Pelt et al. 1994),
although other authors found a value of $540\pm12$ days (Press,
Rybicki, \& Hewitt 1992).  The shorter $\Delta t$ has now been
confirmed (Kundic et al. 1997a, cf.  Serra-Ricart et al. 1999 and
references therein). Since $\Delta t \approx \theta^2 H_0^{-1}$, this
observation allows an estimate of the Hubble parameter. The latest
results for $h$ from 0957+561, using all available data, are $h=0.64
\pm 0.13$ (95\% C.L.) (Kundic et al.  1997a), and $h=0.62\pm0.07$
(Falco et al. 1997), where the error does not include systematic
errors in the assumed form of the lensing mass distribution.

The first quadruple-image quasar system discovered was PG1115+080.
Using a recent series of observations (Schechter et al. 1997), the
time delay between images B and C has been determined to be about
$24\pm3$ days.  A simple model for the lensing galaxy and the nearby
galaxies then leads to $h=0.42\pm0.06$ (Schechter et al. 1997), although
higher values for $h$ are obtained by more sophisticated analyses:
$h=0.60\pm0.17$ (Keeton \& Kochanek 1996), $h=0.52\pm0.14$ (Kundic et
al. 1997b). The results depend on how the lensing galaxy and those in
the compact group of which it is a part are modelled. 

Another quadruple-lens system, B1606+656, leads to $h=0.59 \pm 0.08
\pm 0.15$, where the first error is the 95\% C.L. statistical error,
and the second is the estimated systematic uncertainty (Fassnacht et
al.  1999).  Time delays have also recently been determined for the
Einstein ring system B0218+357, giving $h=0.69^{+0.13}_{-0.19}$ (95\%
C.L.) (Biggs et al. 1999).

Mainly because of the systematic uncertainties in modelling the mass
distribution in the lensing systems, the uncertainty in the $h$
determination by gravitational lens time delays remains rather large.
But it is reassuring that this completely independent method gives
results consistent with the other determinations.

\subsection{Conclusions on $H_0$}

To summarize, relative distance methods favor a value $h\approx
0.6-0.7$. Meanwhile the fundamental physics methods typically lead to
$h \approx 0.4-0.7$.  Among fundamental physics approaches, there has
been important recent progress in measuring $h$ via the
Sunyev-Zel'dovich effect and time delays between different images of
gravitationally lensed quasars, although the uncertainties remain
larger than via relative distance methods.  For the rest of this
review, we will adopt a value of $h=0.65\pm0.08$.  This corresponds to
$t_0= 6.52 h^{-1} {\rm Gyr} = 10 \pm 2$ Gyr for $\Omega_m=1$ ---
probably too low compared to the ages of the oldest globular clusters.
But for $\Omega_m=0.2$ and $\Omega_\Lambda=0$, or alternatively for
$\Omega_m=0.4$ and $\Omega_\Lambda=0.6$, $t_0 = 13\pm2$ Gyr, in
agreement with the globular cluster estimate of $t_0$.  This
is one of the several arguments for low $\Omega_m$, a non-zero
cosmological constant, or both.

\section{Hot Dark Matter Density $\Omega_\nu$}

The recent atmospheric neutrino data from Super-Kamiokande (Fukuda et
al. 1998) provide strong evidence of neutrino oscillations and
therefore of non-zero neutrino mass.  These data imply a lower limit
on the hot dark matter (i.e., light neutrino) contribution to the
cosmological density $\Omega_\nu \gsim 0.001$.  $\Omega_\nu$ is
actually that low, and therefore cosmologically uninteresting, if
$m(\nu_\tau) \gg m(\nu_\mu)$, as is suggested by the hierarchical
pattern of the quark and charged lepton masses.  But if the $\nu_\tau$
and $\nu_\mu$ are nearly degenerate in mass, as suggested by their
strong mixing, then $\Omega_\nu$ could be substantially larger.
Although the Cold + Hot Dark Matter (CHDM) cosmological model with
$h\approx 0.5$, $\Omega_m=1$, and $\Omega_\nu=0.2$ predicts power
spectra of cosmic density and CMB anisotropies that are in excellent
agreement with the data (Primack 1996, Gawiser \& Silk 1998), as we
have just seen the large value measured for the Hubble parameter makes
such $\Omega_m=1$ models dubious.  It remains to be seen whether
including a significant amount of hot dark matter in low-$\Omega_m$
models improves their agreement with data.  Primack \& Gross (1998)
found that the possible improvement of the low-$\Omega_m$ flat (\lcdm)
cosmological models with the addition of light neutrinos appears to be
rather limited, and the maximum amount of hot dark matter decreases
with decreasing $\Omega_m$ (Primack et al. 1995).  For $\Omega_m \lsim
0.4$, Croft, Hu, and Dav\'e (1999) find that $\Omega_\nu \lsim 0.08$.
Fukugita et al. (1999) find more restrictive upper limits with
the constraint that the primordial power spectrum index $n \le 1$, but
this may not be well motivated.

\section{Cosmological Constant $\Lambda$}

The strongest evidence for a positive $\Lambda$ comes from
high-redshift SNe Ia, and independently from a combination of
observations indicating that $\Omega_m \sim 0.3$ together with CMB
data indicating that the universe is nearly flat.  We will discuss
these observations in the next section.  Here we will start by looking
at other constraints on $\Lambda$.

The cosmological effects of a cosmological constant are not difficult
to understand (Felton \& Isaacman 1986; Lahav et al. 1991; Carroll,
Press, \& Turner 1992). In the early universe, the density of energy
and matter is far more important than the $\Lambda$ term on the
r.h.s. of the Friedmann equation. But the average matter density
decreases as the universe expands, and at a rather low redshift ($z
\sim 0.2$ for $\Omega_m=0.3$, $\Omega_\Lambda=0.7$) the $\Lambda$ term
finally becomes dominant.  Around this redshift, the $\Lambda$ term
almost balances the attraction of the matter, and the scale factor $a
\equiv (1+z)^{-1}$ increases very slowly, although it ultimately
starts increasing exponentially as the universe starts inflating under
the influence of the increasingly dominant $\Lambda$ term.  The
existence of a period during which expansion slows while the clock
runs explains why $t_0$ can be greater than for $\Lambda=0$, but this
also shows that there is an increased likelihood of finding galaxies
in the redshift interval when the expansion slowed, and a
correspondingly increased opportunity for lensing by these galaxies of
quasars (which mostly lie at higher redshift $z \gsim 2$).

The observed frequency of such lensed quasars is about what would be
expected in a standard $\Omega=1$, $\Lambda=0$ cosmology, so this data
sets fairly stringent upper limits: $\Omega_\Lambda \leq 0.70$ at 90\%
C.L. (Maoz \& Rix 1993, Kochanek 1993), with more recent data giving
even tighter constraints: $\Omega_\Lambda < 0.66$ at 95\% confidence
if $\Omega_m + \Omega_\Lambda =1$ (Kochanek 1996b).  This limit could
perhaps be weakened if there were (a) significant extinction by dust
in the E/S0 galaxies responsible for the lensing or (b) rapid
evolution of these galaxies, but there is much evidence that these
galaxies have little dust and have evolved only passively for $z \lsim
1$ (Steidel, Dickinson, \& Persson 1994; Lilly et al. 1995; Schade et
al. 1996).  An alternative analysis by Im, Griffiths, \& Ratnatunga
1997 of some of the same optical lensing data considered by Kochanek
1996 leads them to deduce a value $\Omega_\Lambda
=0.64_{-0.26}^{+0.15}$, which is barely consistent with Kochanek's
upper limit.  Malhotra, Rhodes, \& Turner (1997)
presents evidence for extinction of quasars by foreground galaxies
and claims that this weakens the lensing bound to
$\Omega_\Lambda<0.9$, but this is not justified quantitatively.
Maller, Flores, \& Primack (1997) shows
that edge-on disk galaxies can lens quasars very effectively, and
discusses a case in which optical extinction is significant.  But the
radio observations discussed by Falco, Kochanek, \& Munoz (1998),
which give a $2\sigma$ limit $\Omega_\Lambda < 0.73$, are not affected
by extinction.  Recently Chiba and Yoshii (1999) have suggested that
a reanalysis of lensing using new models of the evolution of elliptical
galaxies gives $\Omega_\Lambda=0.7^{+0.1}_{-0.2}$, but Kochanek et
al. (1999, see especially Fig. 4) show that the available evidence 
disfavors the models of Chiba and Yoshii.

A model-dependent constraint appeared to come from simulations of
$\Lambda$CDM (Klypin, Primack, \& Holtzman 1996) and OpenCDM (Jenkins
et al. 1998) COBE-normalized models with $h=0.7$, $\Omega_m=0.3$, and
either $\Omega_\Lambda=0.7$ or, for the open case, $\Omega_\Lambda=0$.
These models have too much power on small scales to be consistent with
observations, unless there is strong scale-dependent antibiasing of
galaxies with respect to dark matter.  However, recent high-resolution
simulations (Klypin et al. 1999) find that merging and destruction of
galaxies in dense environments lead to exactly the sort of
scale-dependent antibiasing needed for agreement with observations for
the \lcdm\ model. Similar results have been found using simulations
plus semi-analytic methods (Benson et al. 1999, but cf. Kauffmann et
al. 1999).

Another constraint on $\Lambda$ from simulations is a claim that the
number of long arcs in clusters is in accord with observations for an
open CDM model with $\Omega_m=0.3$ but an order of magnitude too low
in a \lcdm\ model with the same $\Omega_m$ (Bartelmann et al. 1998).
This apparently occurs because clusters with dense cores form too late
in such models.  This is potentially a powerful constraint, and needs
to be checked and understood.  It is now known that including cluster
galaxies does not alter these results (Meneghetti et al. 1999; Flores,
Maller, \& Primack 1999).

\section{Measuring $\Omega_m$}

The present author, like many theorists, has long regarded the
Einstein-de Sitter ($\Omega_m=1$, $\Lambda=0$) cosmology as the most
attractive one. For one thing, there are only three possible constant
values for $\Omega$ --- 0, 1, and $\infty$ --- of which the only one
that can describe our universe is $\Omega_m=1$.  Also, cosmic
inflation is the only known solution for several otherwise intractable
problems, and all simple inflationary models predict that the universe
is flat, i.e. that $\Omega_m + \Omega_\Lambda=1$.  Since there is no
known physical reason for a non-zero cosmological constant, it was
often said that inflation favors $\Omega=1$.  Of course, theoretical
prejudice is not a reliable guide.  In recent years, many cosmologists
have favored $\Omega_m \sim 0.3$, both because of the $H_0-t_0$
constraints and because cluster and other relatively small-scale
measurements have given low values for $\Omega_m$. (For a summary of
arguments favoring low $\Omega_m \approx 0.2$ and $\Lambda=0$, see
Coles \& Ellis 1997. A review that notes that larger scale
measurements favor higher $\Omega_m$ is Dekel, Burstein, \& White
1997.)  But the most exciting new evidence has come from
cosmological-scale measurements.

{\bf Type Ia Supernovae.}  At present, the most promising techniques
for measuring $\Omega_m$ and $\Omega_\Lambda$ on cosmological scales
use the small-angle anisotropies in the CMB radiation and
high-redshift Type Ia supernovae (SNe Ia).  We will discuss the latter
first.  SNe Ia are the brightest supernovae, and the spread in their
intrinsic brightness appears to be relatively small.  The Supernova
Cosmology Project (Perlmutter et al. 1997a) demonstrated the
feasibility of finding significant numbers of such supernovae.  The
first seven high redshift SNe Ia that they analyzed gave for a flat
universe $\Omega_m=1- \Omega_\Lambda= 0.94^{+0.34}_{-0.28}$, or
equivalently $\Omega_\Lambda= 0.06^{+0.28}_{-0.34}$ ($<0.51$ at the
95\% confidence level) (Perlmutter et al. 1997a).  But adding one
$z=0.83$ SN Ia for which they had good HST data lowered the implied
$\Omega_m$ to $0.6\pm0.2$ in the flat case (Perlmutter et al. 1997b).
Analysis of their larger dataset of 42 high-redshift SNe Ia gives for
the flat cast $\Omega_m = 0.28^{+0.09 +0.05}_{-0.08 -0.04}$ where the
first errors are statistical and the second are identified systematics
(Perlmutter et al. 1999).  The High-Z Supernova team has also searched
successfully for high-redshift supernovae to measure $\Omega_m$
(Garnavich et al. 1997, Riess et al. 1998), and their three HST SNe
Ia, two at $z\approx 0.5$ and one at 0.97, imply $\Omega_m=0.4 \pm
0.3$ in the flat case.  The main concerns about the interpretation of
this data are evolution of the SNe Ia (Drell, Loredo, \& Wasserman
1999) and dimming by dust.  A recent specific supernova evolution
concern that was discussed at this workshop is that the rest frame
rise-times of distant supernovae may be longer than nearby ones (Riess
et al. 1999).  But a direct comparison between nearby supernova and
the SCP distant sample shows that they are rather consistent with each
other (Aldering, Nugent, \& Knop 1999).  Ordinary dust causes
reddening, but hypothetical grey dust would cause much less reddening
and could in principle provide an alternative explanation for the fact
that high-redshift supernovae are observed to be dimmer than expected
in a critical-density cosmology.  It is hard to see why the largest
dust grains, which would be greyer, should preferentially be ejected
by galaxies (Simonsen \& Hannestad 1999).  Such dust, if it exists,
would also absorb starlight and reradiate it at long wavelengths,
where there are other constraints that could, with additional
observations, rule out this scenario (Aguirre \& Haiman 1999).  But
another way of addressing this question is to collect data on
supernovae with redshift $z>1$, where the dust scenario predicts
considerably more dimming than the $\Lambda$ cosmology.
The one $z>1$ supernova currently available, SCP's ``Albinoni''
(SN1998eq) at $z=1.2$, will help, and both the SCP and the High-Z 
group are attempting to
get a few more very high redshift supernovae.

{\bf CMB anisotropies.}  The location of the first Doppler (or
acoustic, or Sakharov) peak at angular wavenumber $l\approx 250$
indicated by the presently available data (Scott, this volume) is
evidence in favor of a flat universe $\Omega_m + \Omega_\Lambda
\approx 1$.  New data from the MAXIMA and BOOMERANG balloon flights
apparently confirms this, and the locations of the second and
possibly third peak appear to be consistent with the predictions (Hu,
Spergel, \& White 1997) of simple cosmic inflation theories.  Further
data should be available in 2001 from the NASA Microwave Anisotropy
Probe satellite.

{\bf Large-scale Measurements.} The comparison of the IRAS redshift
surveys with POTENT and related analyses typically give values for the
parameter $\beta_I \equiv \Omega_m^{0.6}/b_I$ (where $b_I$ is the
biasing parameter for IRAS galaxies), corresponding to $0.3 \lsim
\Omega_m \lsim 3$ (for an assumed $b_I=1.15$).  It is not clear
whether it will be possible to reduce the spread in these values
significantly in the near future --- probably both additional data and
a better understanding of systematic and statistical effects will be
required.  A particularly simple way to deduce a lower limit on
$\Omega_m$ from the POTENT peculiar velocity data was proposed by
Dekel \& Rees (1994), based on the fact that high-velocity outflows
from voids are not expected in low-$\Omega$ models.  Data on just one
nearby void indicates that $\Omega_m \ge 0.3$ at the 97\% C.L.
Stronger constraints are available if we assume that the probability
distribution function (PDF) of the primordial fluctuations was
Gaussian.  Evolution from a Gaussian initial PDF to the non-Gaussian
mass distribution observed today requires considerable gravitational
nonlinearity, i.e. large $\Omega_m$.  The PDF deduced by POTENT from
observed velocities (i.e., the PDF of the mass, if the POTENT
reconstruction is reliable) is far from Gaussian today, with a long
positive-fluctuation tail.  It agrees with a Gaussian initial PDF if
and only if $\Omega_m \sim 1$; $\Omega_m <1$ is rejected at the
$2\sigma$ level, and $\Omega_m \leq 0.3$ is ruled out at $\ge 4\sigma$
(Nusser \& Dekel 1993; cf. Bernardeau et al. 1995).  It would be
interesting to repeat this analysis with newer data.

{\bf Measurements on Scales of a Few Mpc.} A study by the Canadian
Network for Observational Cosmology (CNOC) of 16 clusters at $z\sim
0.3$, mostly chosen from the Einstein Medium Sensitivity Survey (Henry
et al. 1992), was designed to allow a self-contained measurement of
$\Omega_m$ from a field $M/L$ which in turn was deduced from their
measured cluster $M/L$. The result was $\Omega_m=0.19\pm0.06$
(Carlberg et al. 1997).  These data were mainly compared to
standard CDM models, and they appear to exclude $\Omega_m=1$ in such
models.

{\bf Estimates on Galaxy Halo Scales.} Work by Zaritsky et al. (1993)
has confirmed that spiral galaxies have massive halos.  They collected
data on satellites of isolated spiral galaxies, and concluded that the
fact that the relative velocities do not fall off out to a separation
of at least 200 kpc shows that massive halos are the norm.  The
typical rotation velocity of $\sim 200-250$ km s$^{-1}$ implies a mass
within 200 kpc of $\sim 2\times10^{12} M_\odot$.  A careful analysis
taking into account selection effects and satellite orbit
uncertainties concluded that the indicated value of $\Omega_m$ exceeds
0.13 at 90\% confidence (Zaritsky \& White 1994), with preferred
values exceeding 0.3. Newer data suggesting that relative velocities
do not fall off out to a separation of $\sim 400$ kpc (Zaritsky et al.
1997) presumably would raise these $\Omega_m$ estimates.

{\bf Cluster Baryons vs. Big Bang Nucleosynthesis.}  White et al.
(1993) emphasized that X-ray observations of the abundance of baryons
in clusters can be used to determine $\Omega_m$ if clusters are a fair
sample of both baryons and dark matter, as they are expected to be
based on simulations (Evrard, Metzler, \& Navarro 1996). The fair 
sample hypothesis implies that
\begin{equation}
\Omega_m = {\Omega_b \over f_b} = 0.3 \left({\Omega_b \over 0.04}\right)
                                  \left({0.13 \over f_b}\right).
\end{equation}
We can use this to determine $\Omega_m$ using the baryon abundance
$\Omega_b h^2 = 0.019 \pm 0.001$ from the measurement of the deuterium
abundance in high-redshift Lyman limit systems, of which a third has
recently been discovered (Kirkman et al. 1999).  Using X-ray data from an
X-ray flux limited sample of clusters to estimate the baryon fraction
$f_b = 0.075 h^{-3/2}$ (Mohr, Mathiesen, \& Evrard 1999) gives
$\Omega_m = 0.25 h^{-1/2} = 0.3\pm0.1$ using $h=0.65\pm0.08$.
Estimating the baryon fraction using Sunyaev-Zel'dovich measurements
of a sample of 18 clusters gives $f_b = 0.77 h^{-1}$ (Carlstrom et
al. 1999), and implies $\Omega_m = 0.25 h^{-1} = 0.38\pm0.1$.

{\bf Cluster Evolution.}  The dependence of the number of clusters on
redshift can be a useful constraint on theories (e.g., Eke et
al. 1996).  But the cluster data at various redshifts are difficult to
compare properly since they are rather inhomogeneous.  Using just
X-ray temperature data, Eke et al. (1998) conclude that $\Omega_m
\approx 0.45\pm0.2$, with $\Omega_m=1$ strongly disfavored.

{\bf Power Spectrum.} In the context of the \lcdm\ class of models,
two additional constraints are available.  The spectrum shape
parameter $\Gamma \approx \Omega_m h \approx 0.25\pm0.05$, implying
$\Omega_m \approx 0.4\pm0.1$.  A new measurement $\Omega_m = 0.34\pm0.1$
comes from the amplitude of the power spectrum of fluctuations at
redshift $z\sim3$, measured from the Lyman $\alpha$ forest (Weinberg
et al. 1999).  This result is strongly inconsistent with
high-$\Omega_m$ models because they would predict that the
fluctuations grow much more to $z=0$, and thus would be lower at $z=3$
than they are observed to be.

\section{Conclusion}

One of the most striking things about the present era in cosmology is
the remarkable agreement between the values of the cosmological
parameters obtained by different methods --- except possibly for the
quasar lensing data which favors a higher $\Omega_m$ and lower
$\Omega_\Lambda$, and the arc lensing data which favors lower values
of both parameters.  If the results from the new CMB measurements
agree with those from the other methods discussed above, the
cosmological parameters will have been determined to perhaps 10\%, and
cosmologists can turn their attention to the other subjects that I
mentioned at the beginning: origin of the initial fluctuations, the
nature of the dark matter and dark energy, and the formation of
galaxies and large-scale structure.  Cosmologists can also speculate
on the reasons why the cosmological parameters have the values that
they do, but this appears to be the sort of question whose answer may
require a deeper understanding of fundamental physics --- perhaps from
a superstring theory of everything.

\acknowledgments 
I wish to thank St\'ephane Courteau for organizing this exciting
conference!  I am grateful to my students and colleagues for helpful
discussions.  This work was supported in part by NSF and NASA grants
at UCSC.

\end{document}